\documentclass[proof]{WileyASNA-v1}

\articletype{Article Type}%

\received{26 April 2016}
\revised{6 June 2016}
\accepted{6 June 2016}

\begin{document}

\title{White dwarf stars in the big data era \protect
}

\author[1]{Maria Camisassa*}



\authormark{Camisassa 
}

\address[1]{\orgdiv{Departament de Física}, \orgname{Universitat Politècnica de Catalunya}, \orgaddress{ 
\state{Barcelona}, \country{Spain}}}



\corres{* Maria Camisassa \email{maria.camisassa@upc.edu}}

\presentaddress{c/Esteve Terrades 5, 
           08860, Castelldefels, Spain}

\abstract{White dwarf stars are the most common endpoint of stellar evolution. Therefore, these old, numerous and compact objects provide valuable information on the late stages of stellar evolution, the physics of dense plasma and the structure and evolution of our Galaxy. The ESA {\it Gaia} space mission has revolutionized this research field, providing parallaxes and multi-band photometry for
nearly 360\,000 white dwarfs. Furthermore, this data, combined with spectroscopical and spectropolarimetric observations, have provided new information on their chemical abundances and magnetic fields. This large data set has raised new questions on the nature of white dwarfs,
boosting our theoretical efforts for understanding the physics that governs their evolution and for improving the statistical analysis of their collective properties. 
In this article, I summarize the current state of our understanding of the collective properties of white dwarfs, based of detailed theoretical models and population synthesis studies. }




\maketitle


\section{Introduction}\label{sec1}

 White dwarf (WD) stars represent the most common endpoint of stellar evolution, as more than 95\% of main sequence stars will end their lives as WDs. These ancient, compact objects are sustained by electron-degeneracy pressure and undergo a gradual cooling process that extends over several Gyrs ($10^9$ years)  
  \citep[see][for reviews]{2008PASP..120.1043F,2008ARA&A..46..157W,2010A&ARv..18..471A,2016NewAR..72....1G,2019A&ARv..27....7C,2022PhR...988....1S,2022FrASS...9....6I}. This property allows an easy determination of the WD age, making these objects reliable cosmochronometers  for dating stellar populations and their companions in binary systems. Owing to their distinctive properties, WDs are key objects for improving our understanding of the late stages of stellar evolution, as well as planetary systems and the structure and evolution of our Galaxy. Additionally, they provide critical insights into the star formation rate, the initial mass function, the initial-to-final mass relation, and the chemical evolution within the solar neighborhood \citep[e.g.][]{2021MNRAS.505.3165R}. 
  Furthermore, the extreme densities found in WD interiors, which can reach $10^{10}$g/cm$^3$ in the center of the most massive models, make these stars propitious laboratories for studying stellar matter, energy sources and transport under extreme conditions. %

We are witnessing a golden era for stellar astrophysics. Surveys like Sloan Digital Sky Survey \citep[SDSS,][]{2000AJ....120.1579Y}, the Radial Velocity Experiment \citep[RAVE, ][]{2020AJ....160...83S}, the Panoramic Survey Telescope and Rapid Response System \cite[PanSTARRS,][]{2016arXiv161205560C},
among others, are delivering vast datasets of spectra and multi-band photometry for stars across our Galaxy, revolutionizing our knowledge of all phases of stellar evolution \citep[e.g.][]{Kepler2021}. Complementing these efforts, missions like NASA’s Kepler and the Transiting Exoplanet Survey Satellite \citep[TESS,][]{2015JATIS...1a4003R} are providing precise measurements of photometric variations, enabling detailed analyses of stellar variability across different types of stars.

A major step forward has come with the second data release (DR2) from the ESA {\it Gaia} mission. Released in 2018, {\it Gaia} DR2 provided multi-band photometry, proper motions, and parallaxes for 1.3 billion sources \citep{GaiaDR22018}. The highly accurate parallax measurements, combined with photometric data, have enabled the calculation of precise absolute magnitudes for stars, sparking a revolution in all areas of stellar research. {\it Gaia} has been particularly transformative in the study of WD stars \citep[see][for a review]{2024NewAR..9901705T}. With {\it Gaia} DR2, the catalog of WD candidates has expanded to approximately 260,000 \citep{2019MNRAS.482.4570G}, and the volume-limited sample of WDs within 100 pc of the Sun has significantly grown \citep{Jimenez2018,2023MNRAS.518.5106J}. The subsequent release of {\it Gaia} EDR3 in 2020 further expanded the number of high-confidence WD candidates to 360,000 \citep{Fusillo2021}, heralding a new era in the WD research. Additionally, {\it Gaia} has provided around 220 million low-resolution spectra, including approximately 100,000 associated with WDs. 

The precise measurements of WD parallaxes provided by {\it Gaia} have allowed to build local volume-limited samples centered around 20-100\,pc from our Sun. The importance of these samples lies in the fact that they include the older and fainter WDs, that are underrepresented in magnitude-limited samples. Thanks to significant efforts, now we have nearly complete ($\gtrsim 95\%$) volume-limited-samples of WDs: the 100\,pc photometric sample \citep[$\sim 12\,000$ members][]{Jimenez2018,Kilic2020,2023MNRAS.518.5106J}, the 40\,pc spectroscopic sample \citep[$\sim 1\,100$ members][]{2024MNRAS.527.8687O} and the 20pc sample of magnetic WDs \citep[$\sim 200$ members][]{2021MNRAS.507.5902B}. In this article, we summarize the current state of the WD research, placing emphasis on the latest discoveries reached on the basis of these large astronomical datasets. The integration of this extensive data with forthcoming multi-object spectroscopic surveys such as 4MOST, WEAVE, DESI, and SDSS-V is expected to drive continuous advancements in the field of WD research.




\section{Characteristics of the WD population}

\subsection{Mass distribution} \label{md}

\begin{figure}[t]
\centerline{\includegraphics[width=\columnwidth]{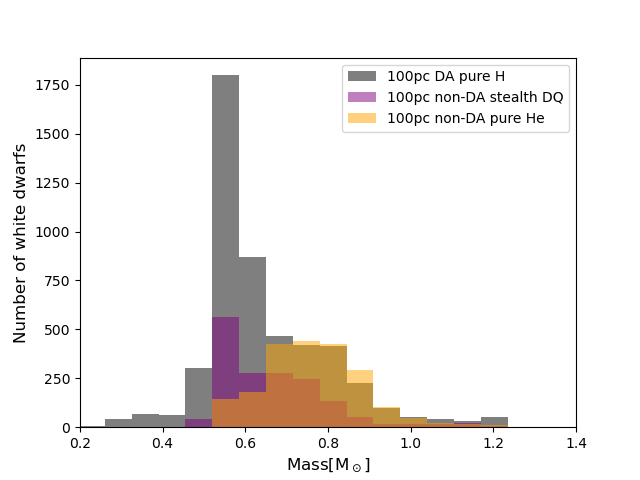}}
\caption{Mass distributions of WDs within 100\,pc with effective temperatures $>6000K$, spectroscopically classified in \cite{2023MNRAS.518.5106J}, obtained via interpolation of the white dwarf models on the {\it Gaia} color-magnitude diagram from \cite{2023A&A...674A.213C} are shown for DA WDs using pure H models (black histogram), and for non-DA WDs using pure He models (orange histogram) and  using stealth-DQ models (purple histogram). Note the main peak around $0.6M_\odot$ and the tail extending towards higher masses in the black histogram. 
\label{mass}}
\end{figure}

The mass distribution of hydrogen(H)-rich WDs from \cite{2023A&A...674A.213C} can be seen in the black histogram of Fig. \ref{mass}. It exhibits a main peak around 0.6 $M_\odot$, with tails extending towards both higher and lower mass, in agreement with different mass distributions along the literature \citep{2015MNRAS.452.1637R, 
2018MNRAS.480.3942H,Kilic2020,2024MNRAS.527.8687O}. This main peak corresponds to the typical end product of stars with initial masses between $\sim 1-2.5 M_\odot$, which lose most of their mass through stellar winds before ending their lives as WDs. Main sequence stars with masses $\lesssim 1 M_\odot$ take times longer than the age of the Universe to evolve to the WD phase. Therefore, WDs with masses  $\lesssim 0.5 M_\odot$, are less frequent and must be formed through binary evolution, where significant mass loss occurs through interactions. The higher-mass tail is the natural consequence of the initial mass function of main sequence stars. More massive WDs are formed from more massive progenitors, which are less frequent, although an important fraction of massive WDs should have formed as a result of stellar mergers \citep{2020A&A...636A..31T,2020ApJ...891..160C,2022MNRAS.511.5462T}.

\subsection{Internal chemical composition}

The WD chemical profile contains valuable information on the evolution of their progenitor stars. Typically, WDs have a carbon(C)-oxygen(O) core, which comprises $\sim 99\%$ of its mass. This core is surrounded by a helium(He)-rich layer, of up to $10^{-2} M_\odot$, which is surrounded by a H-rich layer, of typically $10^{-4} M_\odot$. The mass of the H layer can be a few orders of magnitude smaller than this value in H-deficient WDs.

WDs with masses $\lesssim 0.5 M_\odot$ are born as a result of binary interactions, and their cores never have reached the He ignition. Therefore, these stars have He cores, surrounded by H rich layers that are, in principle, thicker than in typical average-mass WDs \citep{2013A&A...557A..19A}.

WDs more massive than $\gtrsim 1.05 M_\odot$ are the so-called ultra-massive WDs.  These stars can be formed either in single evolution as the descendants of main sequence stars more massive than $\sim 7.5 M_\odot$ \citep{2024ApJS..270...29L,2010A&A...512A..10S,2021A&A...646A..30A} or in binary evolution as a result of stellar mergers \citep{2021ApJ...906...53S,2022MNRAS.512.2972W,2023ApJ...955L..33S}. There is not a clear consensus in the literature on the core-chemical composition of these stars. Some theoretical simulations predict the C burning either in the super asymptotic giant branch, or in the merger event, thus leading to WDs with O-neon(Ne) cores, while some others predict that C burning is avoided thus resulting in WDs with CO cores. 
Despite there is no clear observational evidence that favours CO or ONe compositions, it is likely that both exist in the Universe. Promising channels to infer the chemical composition in the ultra-massive WDs interiors include Astereoseismology \citep{2019A&A...632A.119C}, the study of magnetic fields \citep{2022MNRAS.516L...1C} and the study of the collective properties of the ultra-massive WD population \citep{2021A&A...649L...7C,2022MNRAS.511.5984F,2024Natur.627..286B}.

\subsection{Surface chemical composition - Spectral type} \label{st}

WDs are typically classified into two main categories based on their atmospheric composition. The first category is characterized by the presence of dominant H lines in their spectra. This group, known as DA WDs, accounts for approximately 70\% of all WDs and is defined by their hydrogen-rich envelope. The second category, known as non-DA WDs, encompasses those WDs without H lines in their spectra. Non-DA WDs are further divided into subcategories based on their spectral features. Those with singly ionized He (HeII) are classified as DO, while those displaying neutral He lines (HeI) fall into the DB category. WDs with spectral traces of C and heavier metals are identified as DQ and DZ, respectively. Finally, those showing no detectable spectral lines are classified as DC, characterized by featureless spectra. Traditionally, WD spectral classification has been done through visual inspection of spectra. However, in recent years, machine learning approaches have increasingly been employed to overcome this task \citep[e.g.][]{2023MNRAS.521..760V,2023A&A...679A.127G}.

Along their lives, WDs undergo spectral evolution \citep[see][for a review]{2024Ap&SS.369...43B}. 
WDs formed with a large H content allow it to float up by gravitational settling, thus forming a H-dominated envelope that will remain through their lives. On the other hand, WDs with low H content undergo a different path. If their H mass is extremely low, gravitational settling will never be able to develop a H-dominated envelope, thus retaining a He-dominated envelope through their lives. If not, they will develop a H-dominated envelope, which will later be dilluted in the underlying He mantle due to convective dillution or convective mixing, turning into He-dominated WDs. The dillution of H in the underlying He mantle is supported by the $\sim 25\%$ increase in the fraction of non-DA WDs at low effective temperatures 
\citep{2019ApJ...882..106G,2020MNRAS.492.3540C,2023MNRAS.518.5106J,2023A&A...677A.159T}. 
Finally, cold WDs with He-dominated envelopes will likely dredge-up C traces to the surface by convection at $\sim 12,000 K$, becoming cold DQs (see section \ref{ABbranch}).

\section{The Gaia color-magnitude diagram}

\begin{figure}[t]
	\centerline{\includegraphics[width=\columnwidth]{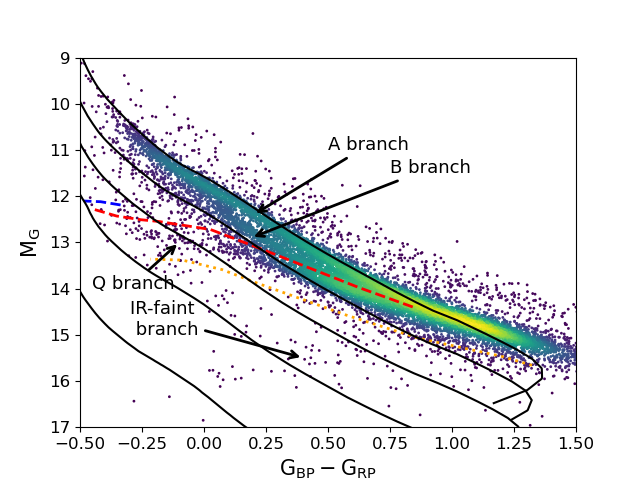}}
	\caption{{\it Gaia} color-magnitude diagram of the WDs within 100\,pc from our Sun, together with H-envelope WD evolutionary models with masses, from top to bottom, 0.53, 0.83, 1.10, 1.29 and 1.369 $M_\odot$ (black solid lines). The  0.53 and 0.83 $M_\odot$ models have CO cores and were calculated in  
 \cite{Camisassa2016}, the 1.10,  1.29 and 1.369 $M_\odot$ have ONe cores and were calculated in  \cite{2019A&A...625A..87C, 2022A&A...668A..58A}. The 1.369 $M_\odot$ model was calculated in  the general relativity framework and represents the most massive ONe-core WD that can be modeled. The crystallization onset for ONe and CO composition and the exsolution onset are shown using blue dashed, red dashed, and orange dotted lines, respectively. This sample contains 12\,263 WDs.  
 \label{gaia}}
\end{figure}

Thanks to the extraordinary astrometric precision achieved by  {\it Gaia}, it has been possible to obtain an exceptionally accurate color-magnitude diagram, clearly separating the WDs from other stars and thus providing valuable information about these objects.
Particularly, {\it Gaia} has revealed the existence of 4 main branches in the WD cooling sequence on the color-magnitude diagram never noticed before. These branches are the A, B, Q and IR-faint branches, which are marked in Fig. \ref{gaia}. This figure shows the WDs within 100\,pc from our Sun in the {\it Gaia} color-magnitude diagram. In this Figure we also plot H-rich (DA) WD evolutionary tracks with masses, from top to bottom, 0.58, 0.66, 0.83, 1.29, and 1.369 $M_\odot$ \citep{Camisassa2016,2019A&A...625A..87C, 2022A&A...668A..58A}.
WDs evolve from bright luminosities and high temperatures, top left corner, to faint luminosities and low temperatures, bottom right corner, as a result of their cooling process. The more massive the WD, the smaller its radius, and therefore the fainter it is. The 1.369 $M_\odot$ evolutionary model was computed in the framework of the general relativity, and has the smallest radius that a WD can have, regardless of whether it harbours a CO or ONe core \citep[see][]{2022A&A...668A..58A,2023MNRAS.523.4492A}. 

At some point in their evolution, WDs cool down enough to start a crystallization process as a result of the Coulomb interactions between the ions, releasing a substantial amount of energy and taking very long periods of time to become entirely crystallized \citep{1968ApJ...151..227V}. The crystallization onset, plotted using blue and red dashed lines for ONe and CO-cores respectively, depends on the mass, occurring at higher temperatures and luminosities for more massive, thus denser, WDs. After crystallization, the main components of the WD interior separate in a process called "exsolution", that also takes very long periods of time and releases energy \citep{2024A&A...683A.101C}. The exsolution onset is depicted using an orange dotted line.

\subsection{The A-B bifurcation}\label{ABbranch}

The A branch is predominantly populated by H-rich (DA) WDs and overlaps a pure H envelope evolutionary model of a WD with mass $\sim 0.6M_\odot$. This overabundance of stars on the A branch is consistent with expectations, as the WD mass distribution prominently peaks around $0.6 M_\odot$ (see Section \ref{md}) and that the WD ages in the solar neighborhood have a nearly uniform distribution. In contrast, the B branch represents a bifurcation from the A branch that occurs when WDs cool to about $\sim 12\,000$K. The B branch is mainly populated by He-rich (non-DA) WDs, and overlaps a pure He envelope evolutionary model of mass $\sim 0.75 M_\odot$.

When employing pure H models to derive photometric masses for DA WDs, we observe that the mass distribution peaks at $\sim 0.6 M_\odot$ (see black histogram in Fig. \ref{mass}), which is consistent with mass distributions determined using different methods. However, when using pure He models for non-DA WDs, the non-DA mass distribution peaks at $\sim 0.75 M_\odot$, largely because most non-DA WDs are located on the B branch \citep[orange histogram in Fig. \ref{mass}, see also][]{Bergeron2019,Fusillo2021,2024MNRAS.527.8687O}
. Nevertheless, a pure He atmosphere may not be the optimal model for cool He-rich WDs with $T_{\rm eff}\lesssim 12\,000$. As discussed in Section \ref{st}, He-envelope WDs are expected to undergo spectral evolution when they reach $T_{\rm eff}\sim 12\,000$, as the outer convective zone penetrates in C-rich layers. This process, known as C-dredge up, has been predicted by theoretical simulations for over four decades and is supported by the observed population of cold DQ stars \citep{1982A&A...116..147K,1986ApJ...307..242P,2005A&A...435..631A,2005ApJ...627..404D,Camisassa2017,BedardDQ,1998MNRAS.296..523M}.
Recent population synthesis studies have shown that He-rich WD models with minor C contamination, below the optical detection threshold, can effectively reproduce the B- branch of the {\it Gaia} bifurcation \citep{2023A&A...674A.213C,2023MNRAS.523.3363B}. These stars, referred to as "stealth DQ" WDs, do not display detectable C features in their optical spectra, but the presence of C in their atmospheres causes continuum absorption that shifts the emission toward bluer wavelengths. This way, "stealth DQ" models with $\sim 0.6M_\odot$ overlap the B branch of the bifurcation \citep[see Fig. 3 in][]{2023A&A...674A.213C}. 
As a consequence, the mass distribution of non-DA WDs, when accounting for this stealth C contamination, peaks at $\sim 0.6 M_\odot$ (purple histogram in Fig. \ref{mass}), aligning with the mass distribution of H-rich WDs and consistent with standard evolutionary formation channels. Additionally, \cite{2023MNRAS.525L.112B} have found evidence of this stealth C contamination in the GALEX UV photometry of WDs on the B branch.

\subsection{The Q branch}\label{Qbranch}

The Q branch is a distinctive, transversal feature in the WD cooling sequence that does not overlap any specific evolutionary track or isochrone. Interestingly, the Q branch coincides with the area where WDs undergo the crystallization process (see Fig. \ref{gaia}).
WDs evolve along their evolutionary track and when they crystallize, they release energy as latent heat and due to a phase separation induced by crystallization \citep[see, e. g.,][]{2019A&A...625A..87C,2022MNRAS.511.5198C,2023ApJ...950..115B}. This energy release slows down the cooling process, thus producing a pile-up of WDs on the Q branch \citep{2019Natur.565..202T}.
However, detailed studies of WDs in the Q branch, using {\it Gaia} kinematics \citep{Cheng2019} and photometry \citep{2021A&A...649L...7C,2024Natur.627..286B}, have revealed that the energy released during crystallization alone is insufficient to fully explain this accumulation, even when accounting for both latent heat and phase separation in the models. Notably, studies of the ultra-massive region of the Q branch pointed a cooling anomaly of nearly 8 Gyrs in a subset of $5-9\%$ of these stars as they pass through the Q branch.
To explain this significant delay in cooling, several studies have proposed the presence of additional energy sources that could substantially slow the cooling process of WDs \citep{2020ApJ...902...93B,2020PhRvD.102h3031H,2021ApJ...919L..12C,2021ApJ...911L...5B, 2021A&A...649L...7C}. The most plausible explanation involves the distillation of $^{22}$Ne during crystallization in WDs with CO cores. In this scenario, a large amount of $^{22}$Ne is transported to central regions of the WD core, releasing significant energy and thereby delaying the cooling process by up to $\sim 10$ Gyr \citep{1991A&A...241L..29I,1996A&A...310..485S,2021ApJ...911L...5B}. Population studies of WD samples in the solar neighborhood \citep{2024Natur.627..286B} and in the globular cluster NGC 6791 \citep{2024A&A...686A.153S} support the occurrence of $^{22}$Ne distillation during crystallization.

\subsection{The IR faint branch}\label{IRfaint}

Finally, the last branch to be identified in {\it Gaia} color-magnitude diagram is a transversal branch of IR-faint WDs located in the 
 faint blue region of the diagram \citep{2022RNAAS...6...36S,Kilic2020}. 
At first glance, these WDs might be mistakenly classified as ultra-massive based solely on {\it Gaia} photometry. However, a closer examination reveals 
that most of these stars 
lie on a WD cooling sequence of mixed-H/He atmosphere WDs, and that the IR-deficit is caused by collision-induced absorption by molecular H due to collisions with helium  
\citep{2022ApJ...934...36B}. It is crucial to note that the atmospheric models for cold mixed H/He WDs are still subject to significant uncertainties \citep{2017ApJ...848...36B}, and current age estimates for some of these stars suggest ages that exceed the age of the Universe. Therefore, further investigation is needed, and observations from the James Webb Space Telescope in the IR spectrum, are expected to provide critical insights into the nature of these stars.

\section{Magnetism in WDs}\label{magneticWD}

\begin{figure}[t]
\centerline{\includegraphics[width=\columnwidth]{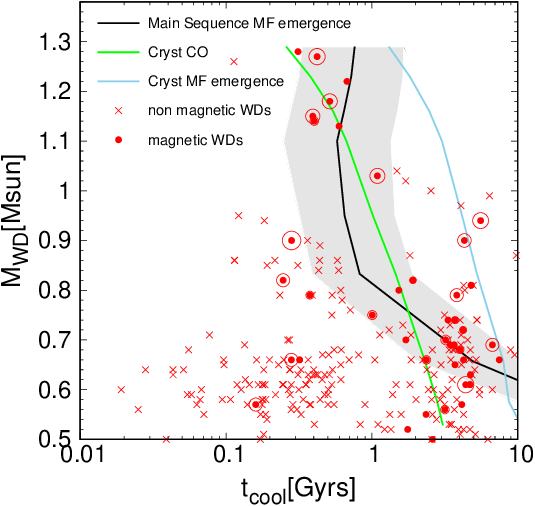}}
\caption{Magnetic field breakout times for CO-core WDs. The crystallization onset is depicted using a green line and the magnetic field breakout time considering a crystallization-driven dynamo is plotted using a blue line. The magnetic breakout times for a main sequence dynamo are shown using a black line, and the shaded are represents the theoretical uncertainties. The magnetic (non-magnetic) WDs from the spectropolarimetric surveys of \cite{2021MNRAS.507.5902B,2022ApJ...935L..12B} are plotted using filled circles (crosses), with a surrounding circle whose radius is proportional to the logarithm of the magnetic field strength. These spectropolarimetric surveys are volume-limited up to 20pc for all WDs and up to 40 pc for those WDs younger than 0.6 Gyrs. Figure adapted from \cite{2024A&A...691L..21C}.
\label{magnetic}}
\end{figure}

Since the discovery of the first magnetic WDs over 50 years ago \citep{1970ApJ...161L..77K,1970ApJ...160L.147A}, the known population has grown significantly, now numbering over 600 \citep[see][for reviews]{2015SSRv..191..111F,2020AdSpR..66.1025F}. 
Magnetic fields have been detected in a wide range of WD classes, including both single and binary systems, across different spectral types, masses, and ages, indicating that magnetism can manifest in nearly any kind of WD \citep{2020AdSpR..66.1025F}. Remarkably, even radio pulsing WDs have been observed in binaries \citep{2023NatAs...7..931P,2016Natur.537..374M}. 
The magnetic field strength in WDs spans a wide range, from 10$^3$ to 10$^9$ G, with no apparent bias in distribution.
Interestingly, there is no clear correlation between magnetic field strength and rotation period. One notable characteristic of magnetic WDs is their tendency to be more massive than their non-magnetic counterparts \citep{2020AdSpR..66.1025F,2020IAUS..357...60K}.

Despite the extensive research on magnetic WDs, both observational and theoretical, the origin of magnetism in WDs remains a topic of active debate. One long-standing hypothesis suggests that magnetic WDs inherit their magnetic fields from their progenitor stars. More recently, another hypothesis proposed that a dynamo-generated magnetic field could arise from the mixing instabilities induced by crystallization in fast-rotating WDs \citep{2017ApJ...836L..28I}. Additionally, interactions in close binary systems have been suggested as possible mechanisms for magnetic field generation, either through a dynamo process during a post-common-envelope phase \citep{2008MNRAS.387..897T} or during a merger event \citep{2012ApJ...749...25G}.

A significant advancement in understanding magnetism in WDs comes from the recent analysis of a 20 pc volume-limited sample of magnetic WDs \citep{2021MNRAS.507.5902B,2007ApJ...654..499K}. These authors meticulously examined each WDs within 20 pc of the Sun for magnetic fields. 
Based on their analysis, \cite{2021MNRAS.507.5902B,2022ApJ...935L..12B} concluded that the incidence of magnetism is notably higher in WDs that have undergone core crystallization compared to those with fully liquid cores. In Fig. \ref{magnetic}, the sample from \cite{2021MNRAS.507.5902B,2022ApJ...935L..12B} is plotted, together with the onset of crystallization indicated using a green solid line; WDs to the left of this line have fully liquid cores, while those to the right have partially or fully crystallized cores. These observations reveal that magnetism is present in about 10\% of non-crystallized WDs, while this fraction increases to approximately 30\% in crystallizing WDs.

The potential for convective motions, induced by crystallization in the overlying liquid mantle, to generate magnetic fields strong enough to match the observed surface fields in WDs has been explored in several studies \citep{2022MNRAS.514.4111G,2024ApJ...961..197M,2024ApJ...969...10C,2024ApJ...964L..15F,2022MNRAS.516L...1C}. However, even if crystallization-driven dynamos can produce the necessary magnetic field strength, this field would not emerge at the surface immediately \citep{2024MNRAS.528.3153B}. In Fig. \ref{magnetic}, we plot our estimation for the magnetic field breakout times \footnote{The magnetic breakout time is the moment when we expect the magnetic field, initially trapped in the WD interior, to emerge to the surface.} 
for crystallization-driven dynamos using a blue solid line \citep{2024A&A...691L..21C}. These long breakout times suggest that crystallization-driven dynamos cannot account for the magnetic fields observed in most stars.

An alternative explanation for the origin of magnetism in WDs is studied in \cite{2024A&A...691L..21C}. WDs with masses $\gtrsim 0.55M_\odot$ are the descendants of main sequence stars with convective cores capable of generating strong dynamo magnetic fields. This hypothesis is supported by magnetohydrodynamical simulations of main sequence convective cores \citep{2009ApJ...705.1000F} and by asteroseismic evidence of strong magnetic fields buried within the interiors of red giant branch stars \citep{2016Natur.529..364S}. The breakout times of these main sequence dynamo magnetic fields are shown in Fig. \ref{magnetic} with a solid black line, where the shaded are allows for theoretical uncertainties. Although these breakout times are subject to enormous uncertainties, the main sequence dynamo emergence hypothesis offers a more plausible explanation for the presence of magnetic fields in WDs. Additionally, it aligns with the well-established fact that magnetic WDs tend to be more massive than their non-magnetic counterparts.

\section{Concluding remarks}\label{Conclusions}

Recent advancements in WD observations have transformed our understanding of these stellar remnants. Particularly, the achievement of nearly complete volume-limited samples of WDs, exempt from most of the observational biases present in magnitude-limited samples has enabled a proper analysis of the characteristics of the WD population. Indeed, the analysis of the A-B bifurcation of the WD cooling sequence in the  100\,pc {\it Gaia} color-magnitude diagram, together with spectroscopic information of these stars, have confirmed 40 years of work in theoretical predictions of WD spectral evolution. Meanwhile, the Q branch has confirmed the theory of WD crystallization, predicted more than 50 years ago \citep{1968ApJ...151..227V}. Although the occurrence of WD crystallization has already been proved in the open cluster NGC 6791 \citep{2010Natur.465..194G}, the quality of the {\it Gaia} observations is significantly better. Furthermore, photometric and kinematic analysis of the Q branch have suggested that a distillation of $^{22}$Ne must take place during crystallization in a selected portion of the WDs \citep{2021ApJ...911L...5B}. The recent analysis of a volume-limited sample of magnetic WDs has provided new insights on the origin of magnetic fields in these stars, showing an increase in magnetic incidence over time. Current possible explanations for this increase invoke a dynamo acting during WD crystallization and the breakout of a magnetic field generated on the main sequence and trapped in the stellar interior for long periods of time \citep{2024A&A...691L..21C}.

In order to analyze the large observational data available, new efforts in the theoretical modeling of WDs have been made. These improvements comprise the inclusion of general relativity effects, the process of $^{22}$Ne distillation upon crystallization and the exsolution process, the modeling of WD magnetic fields and improvements in the atmosphere models, to cite some examples. Moreover, efforts have been placed on performing a proper statistical analysis of the complete WD sample, including machine learning approaches. In the following years we expect a burst of information to be delivered by forthcoming multi-object spectroscopic surveys such as 4MOST, WEAVE, DESI, and SDSS-V, and new observations of  stellar variability will be provided by PLATO space mission. Therefore, we expect new questions to be raised and newer efforts to come to further improve the theoretical modeling of WDs to answer these questions. With this in mind, it is important to recall that WDs are the final stage of the majority of stars, and thus a better understanding of the WD population can help to shed light on enigmas in a wide variety of stars.


\section*{Acknowledgments}

This work is supported by grant RYC2021-032721-I, funded by MCIN/AEI/10.13039/501100011033 and by the European Union NextGenerationEU/PRTR,
the AGAUR/Generalitat de Catalunya grant SGR-386/2021 and by the Spanish MINECO grant PID2020-117252GB-I00.
MC acknowledges the organizers of XMM-Newton 2024 Science Workshop
The X-ray Mysteries of Neutron Stars and White Dwarfs, and A. C{\'o}rsico, L. Athaus, S. Torres and R. Raddi for reading the manuscript and providing their feedback.





\subsection*{Conflict of interest}

The author declares no potential conflict of interests.



\bibliography{Wiley-ASNA}%


\end{document}